\documentclass{article}
\pdfoutput=1
\usepackage{arxiv}

\usepackage[utf8]{inputenc} % allow utf-8 input
\usepackage[T1]{fontenc}    % use 8-bit T1 fonts
\usepackage{hyperref}       % hyperlinks
\usepackage{url}            % simple URL typesetting
\usepackage{booktabs}       % professional-quality tables
\usepackage{amsfonts}       % blackboard math symbols
\usepackage{nicefrac}       % compact symbols for 1/2, etc.
\usepackage{microtype}      % microtypography
\usepackage{graphicx}
\usepackage{doi}
\usepackage{amsmath}
\usepackage{tabularx}
\usepackage[ruled,vlined]{algorithm2e}

\title{Heading perception and the structure of the optic acceleration field}

\author{ \href{https://orcid.org/0000-0002-9286-2091}{\includegraphics[scale=0.06]{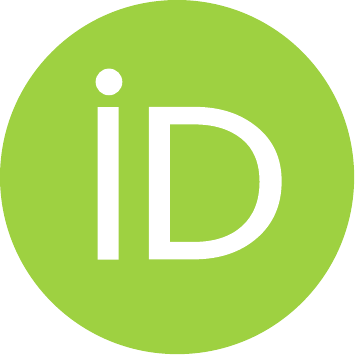}\hspace{1mm}Charlie S. Burlingham} \\
	Department of Psychology\\
	New York University\\
	\texttt{charlie.burlingham@nyu.edu} \\
	\And
	\href{https://orcid.org/0000-0003-0203-1356}{\includegraphics[scale=0.06]{orcid.pdf}\hspace{1mm}Mengjian Hua} \\
	Department of Psychology\\
	New York University\\
	\texttt{mh5113@nyu.edu} \\
	\And
	\href{https://orcid.org/0000-0002-5439-6085}{\includegraphics[scale=0.06]{orcid.pdf}\hspace{1mm}Oliver Xu} \\
	Department of Psychology\\
	New York University\\
	\texttt{yx1797@nyu.edu} \\
	\And
	\href{https://orcid.org/0000-0002-9210-8275}{\includegraphics[scale=0.06]{orcid.pdf}\hspace{1mm}Kathryn Bonnen} \\
	School of Optometry\\
	Indiana University\\
	\texttt{kbonnen@iu.edu} \\
	\And
	\href{https://orcid.org/0000-0002-3282-9898}{\includegraphics[scale=0.06]{orcid.pdf}\hspace{1mm}David Heeger} \\
	Department of Psychology\\
	Center for Neural Science\\
	New York University\\
	\texttt{david.heeger@nyu.edu} \\
}

\renewcommand{\shorttitle}{\textit{arXiv} Template}

\hypersetup{
pdftitle={Heading perception and the structure of the optic acceleration field},
pdfsubject={q-bio.nc, cs.cv},
pdfauthor={Charlie S.~Burlingham, Mengjian ~Hua, Oliver ~Xu, Kathryn ~Bonnen, David J.~Heeger},
pdfkeywords={heading perception, ego-motion, camera motion estimation, optic flow, optic acceleration, time-varying optic flow},
}

\begin{document}
\maketitle

\begin{abstract}
	Visual estimation of heading in the human brain is widely believed to be based on instantaneous optic flow, the velocity of retinal image motion. However, we previously found that humans are unable to use instantaneous optic flow to accurately estimate heading and require time-varying optic flow (Burlingham and Heeger, 2020). We proposed the hypothesis that heading perception is computed from optic acceleration, the temporal derivative of optic flow, based on the observation that heading is aligned perfectly with a point on the retina with zero optic acceleration. However, this result was derived for a specific scenario used in our experiments, when retinal heading and rotational velocity are constant over time. We previously speculated that as the change over time in heading or rotation increases, the bias of the estimator would increase proportionally, based on the idea that our derived case would approximate what happens in a small interval of time (when heading and rotation are nearly constant). In this technical report, we characterize the properties of the optic acceleration field and derive the bias of this estimator for the more common case of a fixating observer, i.e., one that moves while counter-rotating their eyes to stabilize an object on the fovea. For a fixating observer tracking a point on a fronto-parallel plane, there are in fact two singularities of optic acceleration: one that is always at fixation (due to image stabilization) and a second whose bias scales inversely with heading, inconsistent with human behavior. For movement parallel to a ground plane, there is only one singularity of optic acceleration (at the fixation point), which is uninformative about heading. We conclude that the singularity of optic acceleration is not an accurate estimator of heading under natural conditions.
\end{abstract}

% keywords can be removed

\twocolumn

\section{Introduction}
The consensus of opinion in perceptual psychology is that heading, the instantaneous direction of ego-translation in retinal coordinates, is computed from instantaneous optic flow, the velocity of retinal image motion \cite{van1993perception,Lappe1999PerceptionOS,LiSweetStone2006,Sauer2022,Matthis2022}. That is, heading is thought to be estimated from instantaneous optic flow, and these heading estimates are accumulated over time to guide locomotion. Likewise, in computer vision, nearly all camera motion estimation methods explicitly rely on instantaneous optic flow, i.e., they take two video frames or one optic flow field as input. Optic flow evolves over time, however, and this evolution contains information about heading \cite{Burlingham356758,Rieger1983InformationIO,Bandyopadhyay1985PerceptionOR,Wohn19863DMR,Arnspang1988OpticA,Arnspang1990DirectDO,Arnspang1995EstimatingTT,Barron1995MotionAS}. A recent psychophysical study of human perception demonstrated that instantaneous optic flow is insufficient for accurate heading estimation, and that time-varying evolution of optic flow is necessary \cite{Burlingham356758}. Based on this experimental finding, we proposed the "optic acceleration hypothesis" — that the brain computes heading from optic acceleration, the temporal derivative of optic flow. We showed that a heading estimator based on the singularity of the optic acceleration field is unbiased for constant heading and rotation, and speculated that the bias of this estimator would scale with the magnitude of time-varying changes in heading or rotation. 

Here, we follow up on this speculation, characterizing the structure of the optic acceleration field and the bias of the proposed estimator for natural gaze behavior during movement. We simulated a moving retinal observer that fixates on an object (a wall) in its environment, stabilizing its image on the retina. This "fixating observer" generates visual rotations which lead to a radial optic flow pattern centered at the fovea of the eye \cite{Grigo1999DynamicalUO,Lappe1999PerceptionOS,Matthis2022}. Our simulations revealed that optic acceleration in this case has two singularities, one of which is at fixation (when fixating at the plane), and the other of which provides a biased estimate of heading. We characterized this bias as a function of heading, fixation depth, and object depth. We conclude that the second singularity of optic acceleration, considered as a heading estimator, is inconsistent with human behavior.

\section{Optic acceleration}
We first review the mathematical formalisms that relate image motion, observer ego-motion, the projective geometry of the eye, and the depth structure of the environment \cite{Heeger2004SubspaceMF}. Then we review the relevant viewing and movement geometry for the case of a fixating observer.

Optic acceleration is defined as the temporal derivative of optic flow ($\mathbf{u}$) \cite{Heeger2004SubspaceMF}:
\begin{equation}\label{optic_flow}
    \mathbf{u} = p\mathbf{AT}+\mathbf{B\Omega}
\end{equation}
\begin{equation}\label{optic_accel}
    \frac{\partial \mathbf{u}}{\partial t} = \frac{\partial p}{\partial t} \mathbf{AT} + p\mathbf{A} \frac{\partial \mathbf{T}}{\partial t} + \mathbf{B} \frac{\partial \mathbf{\Omega}}{\partial t}
\end{equation}

where $p = \frac{1}{Z(x,y)}$ and
\begin{equation}
\mathbf{T} = \begin{bmatrix} 
    T_x \\ 
    T_y \\ 
    T_z
    \end{bmatrix},
    \end{equation}
    
    \begin{equation}
  \mathbf{\Omega} = \begin{bmatrix} 
    \Omega_x \\ 
    \Omega_y \\ 
    \Omega_z
    \end{bmatrix},
    \end{equation}
    
    \begin{equation}
    \mathbf{A} = \begin{bmatrix} 
    -f & 0 & x \\ 
    0 & -f & y
    \end{bmatrix},
    \end{equation}
    
    \begin{equation}
    \mathbf{B} = \begin{bmatrix} 
    \frac{xy}{f} & -(f+\frac{x^2}{f}) & y \\ 
    f+\frac{y^2}{f} & -\frac{xy}{f} & -x
    \end{bmatrix},
    \end{equation}
    
and $\mathbf{T}$ is the instantaneous translation of the camera in retino-centric coordinate, $\mathbf{\Omega}$ is the instantaneous angular velocity of the camera, $f$ is the focal length of the camera, $x$ and $y$ are coordinates in the image plane (retina), and $Z(x,y)$ is the depth map, where $Z$ is an axis emerging from the pupil out into the world.  
    
For the case of motion along a circular path, in which the line of sight (gaze direction) is yoked to the direction of translation (e.g., while looking at a smudge on the windshield momentarily while on a roundabout), the translational and rotational acceleration are zero. Thus, equation \eqref{optic_accel} simplifies to 

\begin{equation}\label{pnas_eq}
\frac{\partial \mathbf{u}}{\partial t} = \frac{\partial p}{\partial t} \mathbf{AT}.
\end{equation}

The singularity of optic acceleration is the image location for which acceleration is zero. To find this point, we set the equation equal to zero and find that the singularity location is equal to the retinal angle of the heading, if $\frac{\partial p}{\partial t}$ is non-zero (which it will be whenever one is moving):

\begin{equation}
\\ \begin{bmatrix} h_x \\ h_y \end{bmatrix} := \tan^{-1} \left( \frac{1}{T_Z} \begin{bmatrix} T_X \\ T_Y \end{bmatrix} \right)  = \tan^{-1} \left( \frac{1}{f} \begin{bmatrix} x_0 \\ y_0 \end{bmatrix} \right) 
\end{equation}

This result was first shown by Barron \cite{Barron1995MotionAS} and later expressed more compactly and explored further by our group. We proposed the optic acceleration hypothesis \cite{Burlingham356758}: for constant heading and rotation, the singularity of optic acceleration provides an unbiased estimate of heading. We speculated that for small changes in heading or rotation over time, the bias of the estimator would scale monotonically. Even though constant heading and rotation is a case that occurs rarely in real world vision, we speculated that the  optic acceleration singularity may still provide a reasonably good estimate of heading for more common gaze behavior, when rotation and heading change over time. 

For the case of a motion on a circular path, the optic acceleration field has another interesting property: its singularity $x_0$ does not depend on depth $Z$, so while the optic acceleration vectors are scaled by the inverse depth $p$, their orientations are invariant to depth. This has not been observed previously.

\begin{figure}
	\centering
\includegraphics[width = 0.4 \textwidth]{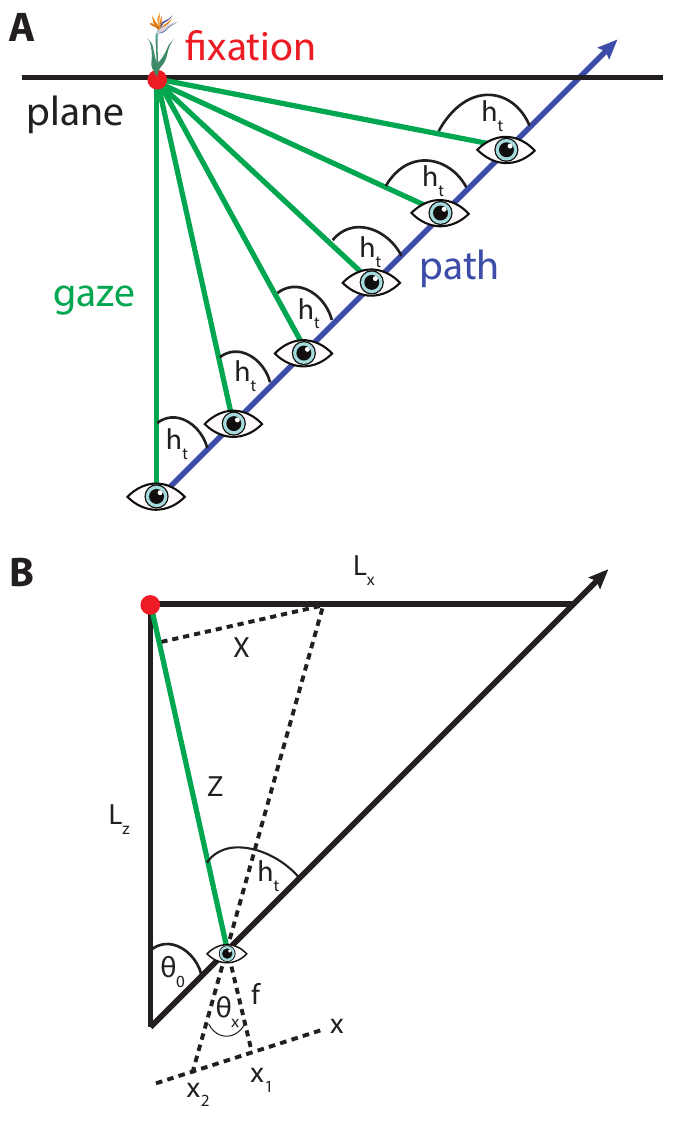}
	\caption{Fig. 1. Movement and viewing geometry of a fixating observer. From a bird's eye view. \textbf{A.} A retinal observer (eyeball icon) traverses a straight path (blue vector), while fixating on a point (red dot and flower) on a plane (black line). The gaze angle (green line) changes over time, generating visual rotations and changes in the heading angle $h_t$ over time. \textbf{B.} Diagram replotted from A. and overlaid with projective geometry and mathematical notation used in in-text equations. $x_1$ and $x_2$ denote the image locations of the two singularities in the optic acceleration field (the first is at fixation and the second is offset from the heading), $f$ denotes the focal length of the projection, $x$ denotes the horizontal axis of the image coordinate system, and $Z$ and $X$ denote the axes of the viewer-centered coordinate system. $L_x$, $L_z$, and $\theta_0$ are a function of the path and fixation point. Perspective projection is assumed in the diagram, but the angle of the second singularity $\theta_x$ is invariant to the projection method, i.e., will generalize to a hemispherical retinal projection.}
	\label{fig:fig1}
\end{figure}

Here, we instead consider a natural scenario of an observer fixating on a point in their environment while travelling along a straight path. We assume that the observer does not have vertical motion or non-yaw rotation, which implies that $\Omega_x$, $\Omega_z$, and $T_y$ are zero (note, however, that our conclusions don't depend on this assumption).
Therefore, it follows that: 

\begin{equation}
\begin{aligned}
     \frac{\partial \mathbf{u}}{\partial t} = & - \frac{Z'}{Z^2} 
     \begin{bmatrix} 
    -f & 0 & x \\ 
    0 & -f & y
    \end{bmatrix} 
    \begin{bmatrix} 
    T_x \\ 
    0 \\ 
    T_z
    \end{bmatrix} \\ &
    + \frac{1}{Z} \begin{bmatrix} 
    -f & 0 & x \\ 
    0 & -f & y
    \end{bmatrix} 
    \begin{bmatrix} 
    \frac{\partial T_x}{\partial t} \\ 
    0 \\ 
    \frac{\partial T_z}{\partial t}
    \end{bmatrix}\\ & + \begin{bmatrix} 
    \frac{xy}{f} & -(f+\frac{x^2}{f}) & y \\ 
    f+\frac{y^2}{f} & -\frac{xy}{f} & -x
    \end{bmatrix} \begin{bmatrix} 
    0 \\ 
    \frac{\partial \Omega_y}{\partial t} \\ 
    0
    \end{bmatrix}.
\end{aligned}
\end{equation}

We assume that the singularity is on the horizontal axis $y = 0$ on the image plane such that:

\begin{equation}\label{singularitieseq}
\begin{aligned}
    0 = & -\frac{Z'}{Z^2} (xT_z - f T_x) \\& + \frac{1}{Z} (x \frac{\partial T_z}{\partial t} - f \frac{\partial T_x}{\partial t}) \\& - \frac{\partial \Omega_y}{\partial t} (f+\frac{x^2}{f}).
\end{aligned}
\end{equation}

This resembles a quadratic equation of $x$. However, the first two coefficients contain Z and Z', which themselves depend on $x$ and/or $t$, so it is in fact a more complex equation. 

To simplify further, we must specify the observer's path of travel and environment, which determines the depth map and the observer's translation and rotation. In the case we consider, the environment is a plane, the observer's gaze is initially aligned with the plane normal, their fixation is always locked on a single point on the plane, and their path of travel is a straight line off to the left or right of the gaze direction by some initial angle $\theta_0$ ($\theta_h$). We assume the observer moves at a constant speed $s$. The time-varying depth map Z(x,t) depends on the translation velocity  $\mathbf{T}$, rotational velocity $\Omega_y$, and the 3-D depth structure of the environment (a plane in this case).

For the fixating observer we have specified, we observe that the time-varying angle between the gaze vector and the plane normal is:

\begin{equation}
\rho_t = -\tan^{-1}[ \frac{L_y \tan\theta_0 - (\frac{L_y}{\cos\theta_0} - st)\sin\theta_0} { ( \frac{L_y}{\cos\theta_0} - st) \cos\theta_0 }  ]. 
\end{equation}

The current heading angle is the sum of this and the initial heading:

\begin{equation}
h_t = \theta_0 + \rho_t,
\end{equation}

where $\theta_0 = \tan^{-1}(\frac{L_x}{L_z})$. Therefore, 

\begin{equation}
\begin{aligned}
    T_x & = s \sin h_t \\
    T_z & = s \cos h_t \\
    \Omega_y & = h'\\
    \frac{d\Omega_y}{dt} &= h''\\
\end{aligned}
\end{equation}

We used computational simulations of a retinal fixating observer to characterize the properties of optic flow and optic acceleration in this case. This allowed us to avoid making any approximations, as Eq. \eqref{singularitieseq} proved to be analytically intractable.

\section{Simulation results}

\subsection{Simulation methods}
We simulated a moving, fixating retinal observer using code introduced by Matthis et al. 2022 \cite{Matthis2022}, which returns a retinal optic flow field and relies primarily on the geom3d toolbox in Matlab. The code creates a fronto-parallel plane with adjustable size, depth, and resolution, and simulates retinal image motion for a moving, fixating observer (with a spherical pinhole model of image formation). We made two notable modifications to the original code. (1) To use a fronto-parallel planar environment, we changed the way that objects are projected to the retina (because the original code is a specific to a ground plane environment). (2) We computed the ground truth heading by projecting the instantaneous translation vector of the eye onto the retina. We refer the reader to Matthis et al. 2022 for specific details on the simulation methods and implementation, as our code was otherwise largely the same \cite{Matthis2022}. Because the projection surface of the eye is spherical, to visualize the retinal motion fields in 2 dimensions, we had to choose a particular map projection method. We used a projection method which makes eccentricity in our vector field plots denote great circle distance along the retinal hemisphere, such that degrees of visual angle are easily interpreted as arc lengths \cite{Matthis2022}. We assumed a horizontal and vertical field-of-view (FOV) of 120º, which is slightly less than the field of view of one human eye \cite{Matthis2022}. The translation speed of the observer was 1.5 m/s, the plane depth was varied between 0.5 and 22.5 m, the rotation velocity of the eye (avg. over time) varied from 0 to 5.5 º/s (depending on the heading) for a plane depth of 12.5 m, the movement duration was 0.15 s (for a +30º heading), and the resolution of the interpolated retinal flow/acceleration fields was 2.5 points/deg$^2$. In order to estimate the optic acceleration field, we interpolated the retinal optic flow field at a set of fixed retinal positions (a lattice), and computed a finite difference approximation of the temporal derivative by subtracting subsequent frames of the flow field and dividing by d$t$ (the inter-frame-interval of the simulation in seconds). We also tried using a 5-tap Farid \& Simoncelli derivative filter \cite{Farid_Simoncelli}, but it didn't make a noticeable difference in the simulation (although it does in real data, see Appendix). We estimated the singularities of optic flow and optic acceleration by finding the minima of the absolute value of a horizontal slice of the vector field at y = 0º (horizontal meridian of the image). The singularity not at fixation (x = 0º) was treated as an estimate of the heading. When we varied plane depth in our simulations, we also scaled the width and height of the plane by its depth to keep the (pre-interpolation) retinal resolution of the vector field constant.

\subsection{Singularity of optic flow and optic acceleration are aligned with heading when travelling in the direction of gaze}
First we examined the structure of the optic acceleration field when the observer was travelling in the direction of gaze (i.e., 0º) and fixating on a point on the plane (i.e., there was no eye rotation). In this case, the acceleration field was centered at fixation, and therefore its singularity was aligned with the heading (Fig. 2). The flow field had a similar structure and its singularity was also aligned with the heading, as expected. Optic acceleration was approx. 90 times larger than optic flow on average across the retina.

\begin{figure}
	\centering
\includegraphics[width = 0.4 \textwidth]{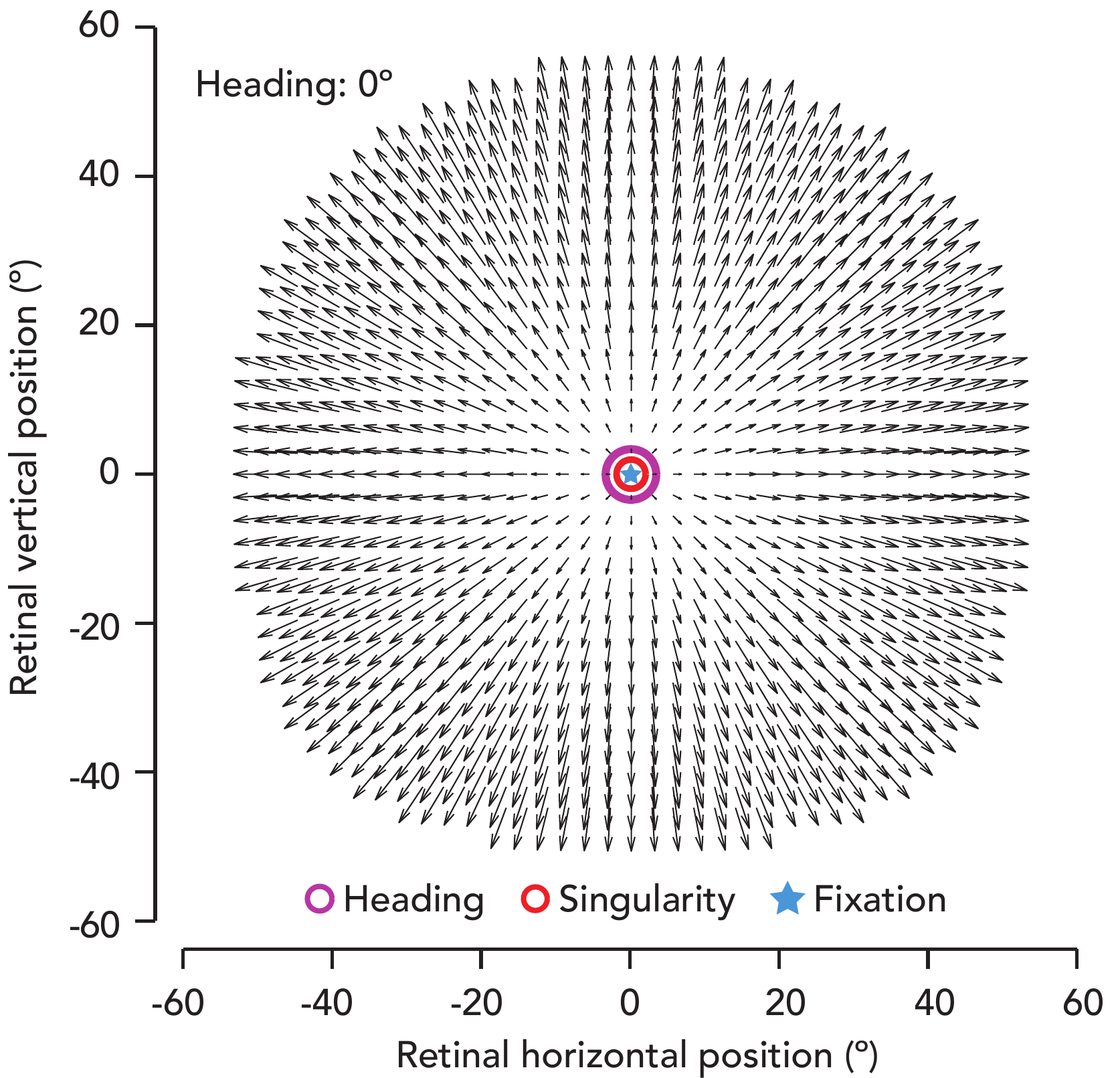}
	\caption{Structure of the optic acceleration field when travelling in the direction of gaze. Black vectors, retinal optic acceleration field for a heading of 0º, i.e., in the direction of gaze. Abscissa and ordinate are retinal location in degrees of visual angle. Red circle, singularity. Purple circle, heading. Vectors auto-scaled for visibility. The flow field has a similar structure in this case, but with longer vectors (not shown).}
	\label{fig:fig2}
\end{figure}

\begin{figure}
	\centering
\includegraphics[width = 0.4 \textwidth]{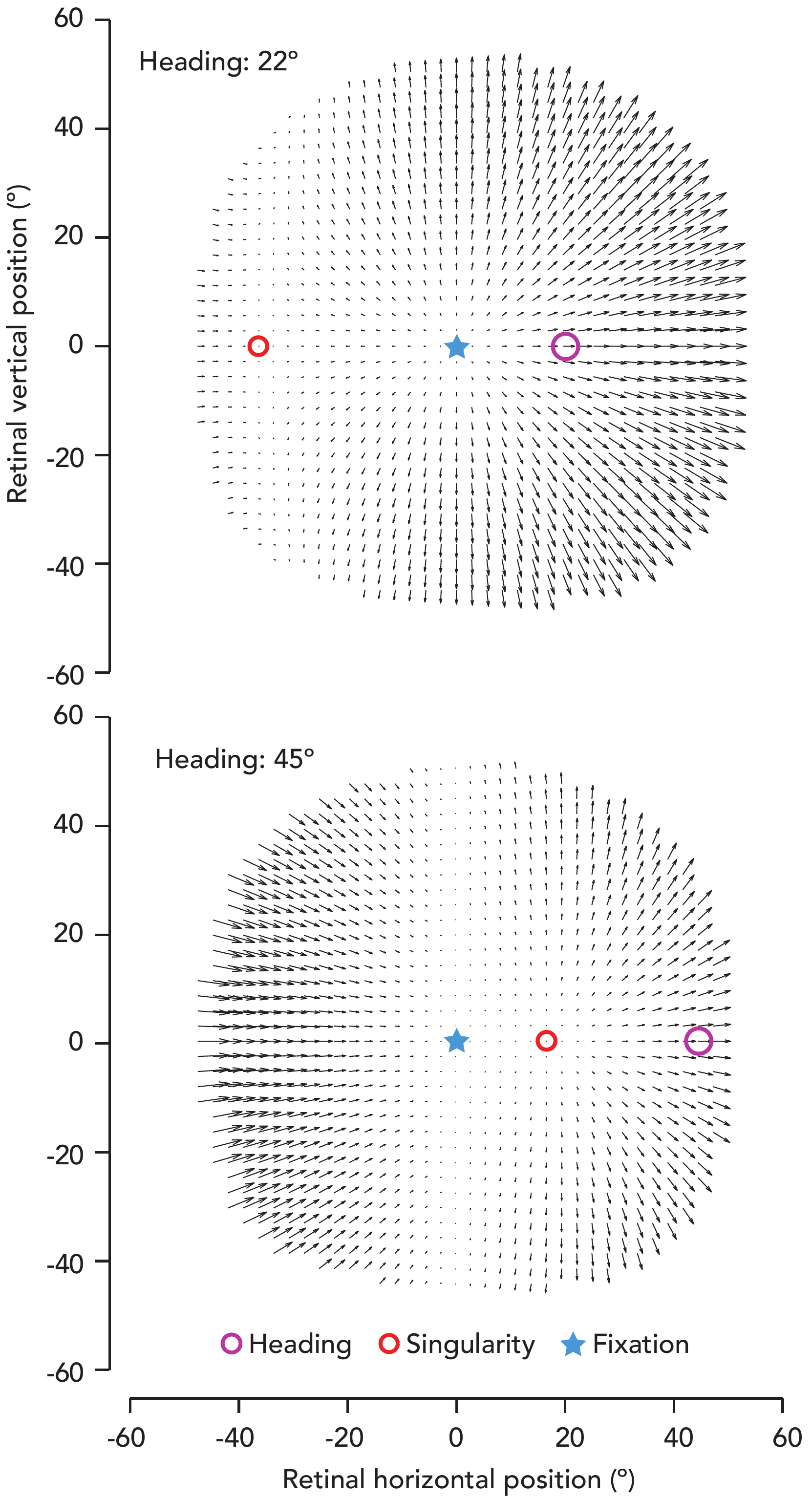}
	\caption{Optic acceleration for a fixating observer has two singularities and its structure depends on heading angle. Same format as Fig. 2. Top panel, heading of +22º. Bottom panel, heading +45º. }
	\label{fig:fig3}
\end{figure}

\begin{figure}
	\centering
\includegraphics[width = 0.41 \textwidth]{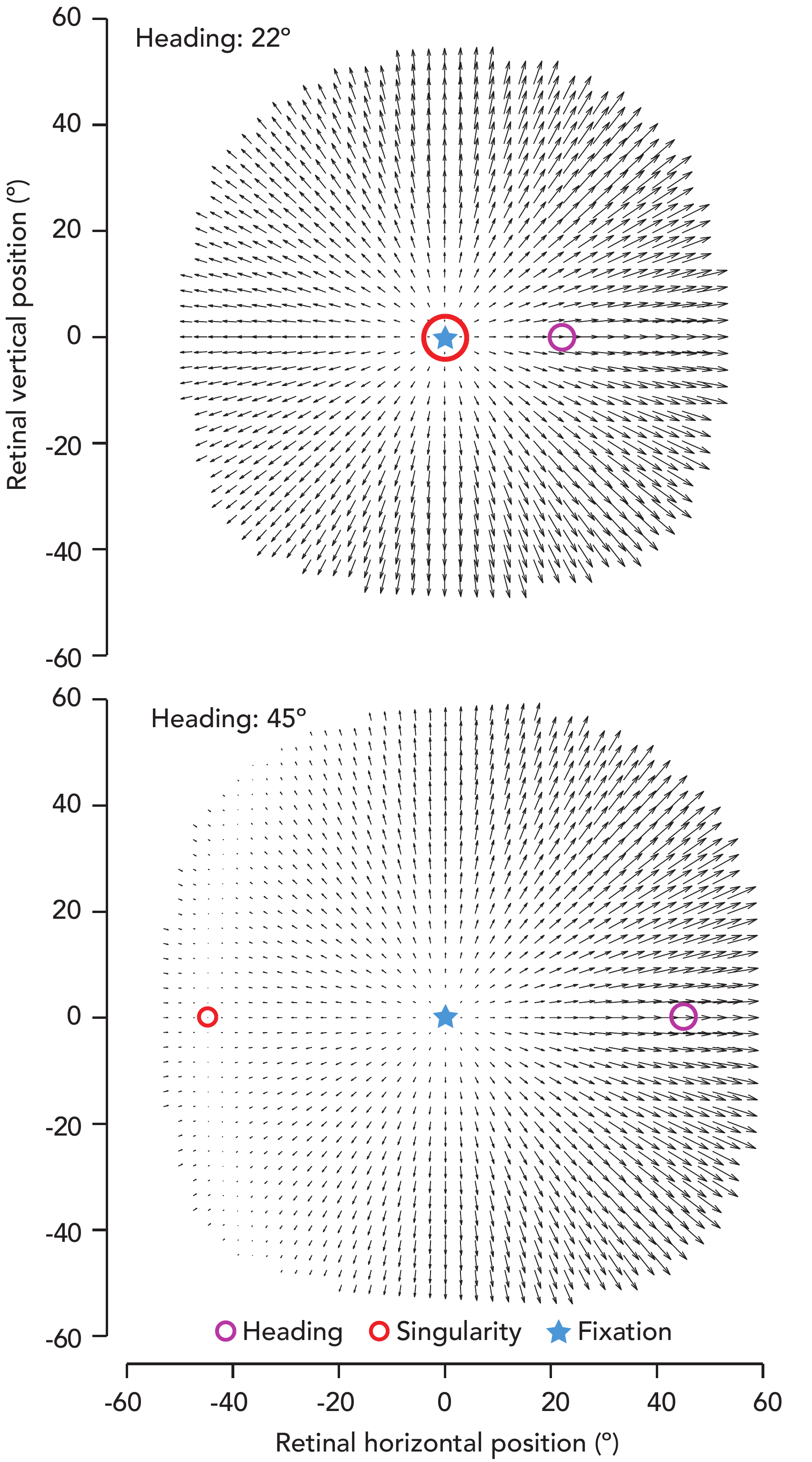}
	\caption{Optic flow for a fixating observer. Same format as Fig. 3, but showing optic flow instead of optic acceleration. Top panel, heading of +22º. Bottom panel, heading +45º. }
	\label{fig:fig4}
\end{figure}

\begin{figure}
	\centering
\includegraphics[width = 0.4 \textwidth]{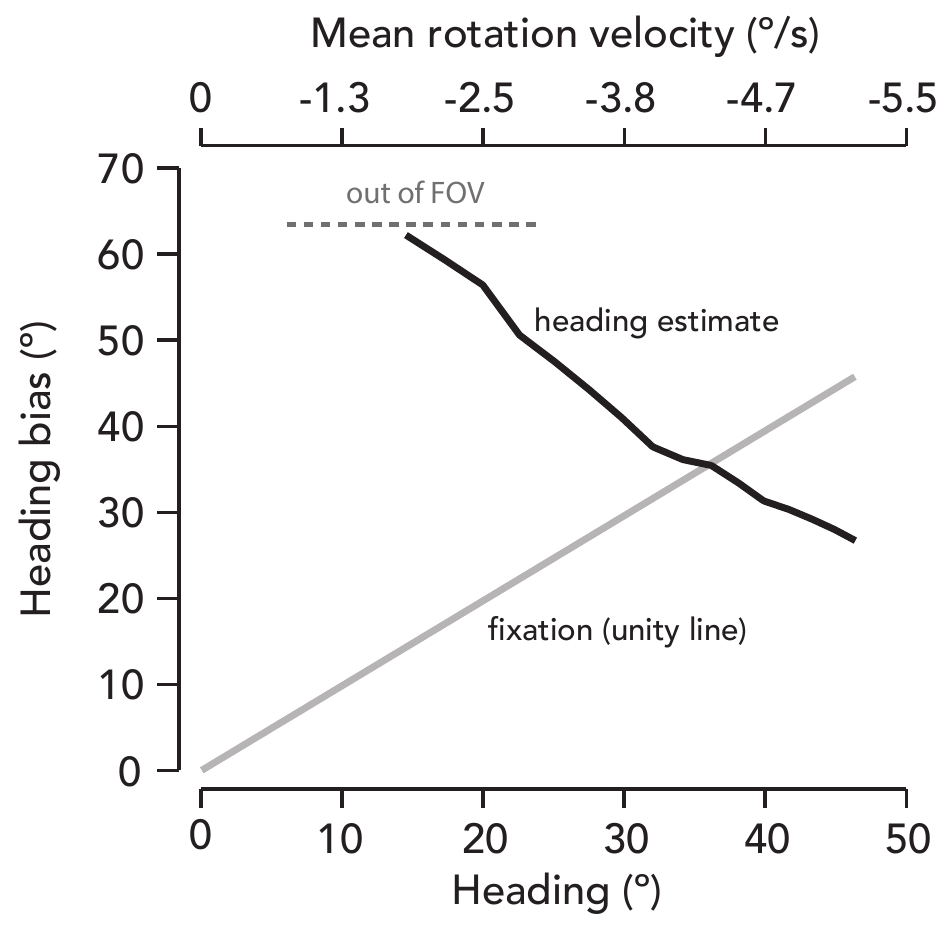}
	\caption{Heading bias scales inversely with heading. Grey line, bias of first singularity of optic acceleration (retinal position minus heading) as a function of heading. The first singularity is always at 0º, fixation, when the observer fixates on the plane, so its bias forms a unity line (plotted as a reference). Black line, bias of second singularity of optic acceleration (i.e., the heading estimate) as a function of heading. Bias is multiplied by -1 for visualization purposes. Black dashed line, second singularity of optic acceleration continues outside the field of view of the simulated eye. Second abscissa, mean rotation velocity of eye over time.}
	\label{fig:fig5}
\end{figure}

\subsection{Heading bias for a fixating observer scales inversely with heading}
Next, we simulated the optic acceleration field when heading was larger, either +22º or +45º. In this case, there were two singularities, one aligned with fixation when the observer fixated on the plane (which makes sense because there was no motion there by definition), and a second to the left of the true heading, offset by some bias (Fig. 3). Depending on the heading angle, the second singularity sometimes took the form of a vertical "trough" extending from the top to the bottom of the retina (Fig. 3, top panel), and in other cases was focal (Fig. 3, bottom panel), like a focus of expansion. Likewise, the flow field also had two singularities, one at fixation and one to the left of heading (Fig. 4), but much further left than the second singularity of optic acceleration.

For the second singularity of optic acceleration, the amount of bias decreased with the absolute heading angle, approaching zero for headings close to 90º (Fig. 5).

\subsection{Heading bias is nearly invariant to fixation depth}
Next, we wanted to understand how the acceleration field for a fixating observer changes with object depth. To test this, we varied the depth of the plane (which was yoked to the fixation depth) from 1 to 10 m from the observer. Surprisingly, the singularity of optic acceleration was nearly invariant to the depth of the plane, showing only a 5º change in position for a 9 m change in depth. We used a heading of 45º in this simulation, but the results were similar for various headings. Specifically, the singularity position increased from 15º to 19º as the fixation depth changed from 1 to 4 m, and then approached an asymptote for depths larger than 4 m. On the other hand, like with optic flow, the magnitude of the acceleration vectors decreased as a function of depth \footnote{For the translational component of optic flow, the scale factor on the vectors is 1/$Z$ \cite{Heeger2004SubspaceMF}, so the combined rotation+translation optic flow field also scales roughly according to the inverse depth (on average across space). The scale factor of the acceleration field is more complicated (see Eq. 2), but still decreases as a function of depth.}.

\begin{figure}
	\centering
\includegraphics[width = 0.48 \textwidth]{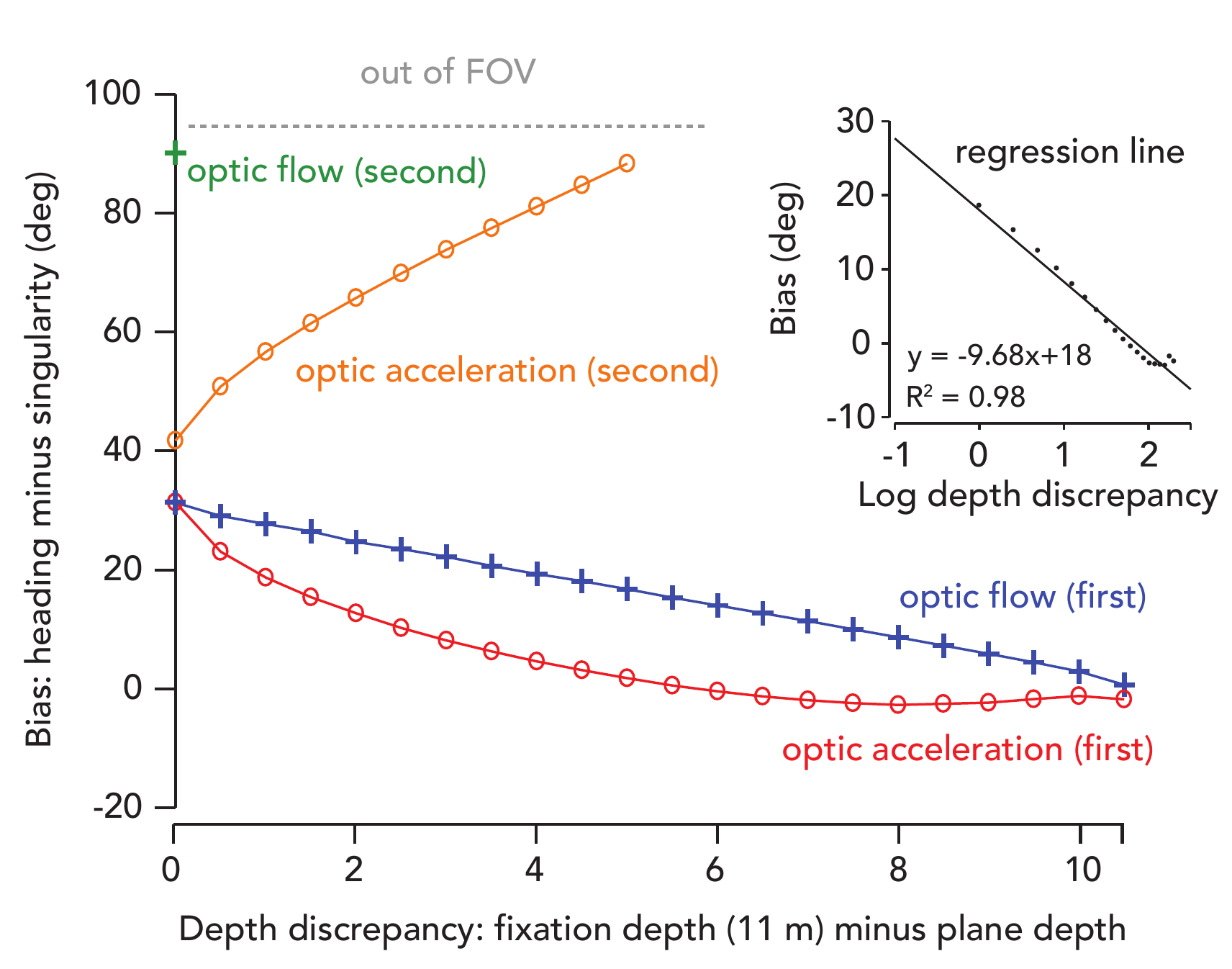}
	\caption{Heading bias varies with depth between fixation and plane. Yellow circles, bias of the second singularity of singularity of optic acceleration as a function of depth discrepancy, i.e., the difference between fixation depth (11 m) and plane depth (variable). So a depth discrepancy of 10 m implies a plane depth of 1 m from the observer. Bias in this figure is actually -1*bias for visualization purposes. The ground truth heading was +30º. Red circles, bias of first singularity of optic acceleration as a function of depth discrepancy. Note that it aligns with fixation for a depth discrepancy of 0 m. Orange circles, bias for second singularity of optic acceleration. Green cirles, bias of second singularity of optic flow. Blue crosses, bias of first singularity of optic flow. Grey dashed lines, bias extrapolated outside of FOV (confirmed to be increasing with separate simulations). Inset shows optic acceleration first singularity bias curve, re-plotted but with a log abscissa, against best-fit regression line.}
	\label{fig:fig6}
\end{figure}

\subsection{Heading bias falls to zero for increasingly close objects}
In real environments there are usually many objects at different depths projecting to the retina. However, in our first simulation, the observer fixated on a fronto-parallel plane as they moved. To understand how optic flow and optic acceleration depend on object depth, we performed a simulation in which the fixation depth and plane depth were decoupled, to mimic typical scenarios in which there is an object in front or behind the fixation point (Fig. 6). 

First, we moved the plane closer than the fixation depth (which we fixed at 11 m). In this case, both the flow field and the acceleration field had two singularities, neither at fixation. Note that there will always be an additional singularity at the fixation point, because the retinal image is stable there. But for the purposes of this simulation, the observer fixated on a infititessimally small point beyond a transparent plane, therefore the fixation singularity was not visible in the retinal motion fields. So in a more general sense, there are three singularities in this scenario: a fixation singularity, a first singularity, and a second singularity. The first and second singularities varied in retinal position depending on heading and "depth discrepancy," i.e., the fixation depth minus the plane depth. We found that the bias of the first singularity of optic flow (i.e., corresponding to fixation for a depth discrepancy of 0 m) approached zero approximately linearly as the depth depth discrepancy increased (i.e., as the plane projecting to the retina came closer to the observer). On the other hand, the bias of the first singularity of optic acceleration fell off approximately logarithmically with depth discrepancy. To test if the falloff was actually logarithmic, we re-plotted the bias curve on a log-linear plot and fit a line to it. The $R^2$ was 0.98 and there was a visible deviation from linearity, indicating that the bias did not fall off logarithmically. Overall, the bias of the first singularity was 10.7º larger for optic flow than optic acceleration (on average across all positive depth discrepancies). 

Even though the first singularity of optic flow and optic acceleration were aligned with the heading when the object was close, we do not consider this singularity as a serious candidate for a heading estimator because it only works when an object is close to you and you are fixating beyond it. The reason the singularity of optic flow is aligned with the heading is trivial and well known — translational, but not rotational optic flow depends on inverse depth (see Eq. \eqref{optic_flow}), so the retinal motion of objects very close to you is dominated by translation, and hence it has a singularity near the heading. For optic acceleration, the argument is nearly the same, but it's because the change in the inverse depth over time is larger when an object is closer to you (see Eq. \eqref{optic_accel}), whereas the translational acceleration of the observer depends on the fixation depth (not the object depth), and so the translation term dominates the equation, leading to a singularity near the heading. The first singularity is also inconsistent with human behavior — people can perceive heading accurately while fixating at a point on a fronto-parallel plane \cite{Grigo1999DynamicalUO}, whereas the first singularity is at fixation in this case and hence completely uninformative. We examine its 'bias' in Fig. 6 only to compare its position directly with that of the second singularity.

The second singularity of optic flow was already large (and outside of the FOV for small headings) and its bias increased even more as a function of depth discrepancy \footnote{We didn't notice this singularity at first because it was so often outside the FOV, being much larger than the the second singularity of optic acceleration. We discovered that this singularity's bias increased as a function with depth (like the second singularity of optic acceleration) via additional simulations using larger headings (e.g., 45º, not shown here).}. The second singularity of optic acceleration also increased as a function of depth discrepancy and exited the FOV after a depth discrepancy of 5.5 m. Overall, the bias of the second singularity was 48.4º larger for flow than acceleration (note: using one point only, depth discrepancy of 0 m). 

Objects are also located at a greater depth than where we are looking in real scenarios. To test how this affects optic flow and acceleration, we repeated the previous simulations, but moved the plane behind the fixation depth (11 m), 0 to 10.5 m away (so up to 22.5 m from the observer). For small negative depth discrepancies, each of the four curves in Fig. 6 seemed to continue, but this pattern fell apart after a depth discrepancy of -5 m, at which point the acceleration field no longer had any singularities and simply resembled a pure rotational flow field. 11 m is a large fixation distance, which could possibly affect these results. To address this, we repeated the same simulation and held the heading and depth discrepancy constant (30º and -5 m), but moved the fixation depth to 2 m and the plane depth to 7 m. We observed the same outcome — an acceleration field with no singularities and a pure rotational structure.

\section{Discussion}
In this technical note we characterized the structure of the optic acceleration field for a fixating observer. We found very large biases for a heading estimator based solely on the singularity of optic acceleration, which decrease as a function of rotation velocity. This sheds doubt on this estimator as a hypothesis for human perception, given that humans achieve low bias (< 2º) under similar conditions, which \textit{increase} as a function of rotation velocity \cite{Grigo1999DynamicalUO}. This doesn't mean that the visual system doesn't use optic acceleration, but just that it is unlikely to estimate heading solely by locating its singularity.

The overarching motivation behind this work was to understand how the visual system determines heading. In the perceptual psychology literature, this question has often been approached in terms of the "rotation problem". That is, if you're looking in the direction you're moving, the optic flow field radiates outward from the heading. In this case, heading perception becomes trivial — you just locate the singularity of the flow field. However, if you rotate your eyes or body while moving, the flow field becomes distorted, such that its singularity is displaced from the heading — hence the "rotation problem." While this formulation is common, we believe it is misleading without further elaboration. To elaborate, one must specify three influences on image formation: (1) the observer's gaze behavior, (2) path of travel, and (3) the depth structure of the environment \cite{Lappe1999PerceptionOS}. 

We start by discussing the first and most critical aspect, gaze behavior. Animals with moving eyes and a fovea make reflexive fixating eye movements as they move, stabilizing the image of their target on the high-resolution photoreceptors in the fovea. This gaze behavior introduces a geometric constraint that links the translational and rotational velocities of the eye, i.e., the eye must counter-rotate at a time-varying velocity in the direction opposite to the heading \cite{Bandopadhay1986VisualNB,Lappe1999PerceptionOS}. We will refer to this as a "fixating observer." For a fixating observer in a realistic scene (i.e., containing objects at different depths), there is always one singularity of optic flow at the center of the retinal image, i.e., the fovea, where the object being stabilized project to. Therefore, this singularity is aligned with heading in only one case for a fixating observer: when the observer walks toward the fixated object. Even this is true only on average over time for natural locomotion, because while walking the body translates along all three axes during the gait cycle, causing the heading to oscillate in retinal location, despite counter-rotations that stabilize the target \cite{Matthis2022}. For all other headings, this singularity solely reflects gaze and gives no information about heading. In addition to this gaze-centered singularity, there are in fact often many more theoretical singularities in the velocity field, each corresponding to different depth planes, a consequence of motion parallax. Note that in reality, some of these singularities are 'missing' or occluded by objects at another depth planes, because real scenes have complex structures. Thus, for a fixating observer in a natural scene, the velocity field is discontinuous, lacks coherent global structure, but always has a singularity centered at the fovea. 

Some studies in perceptual psychology have focused on the gaze behavior of a fixating observer \cite{Grigo1999DynamicalUO,Lappe1999PerceptionOS}, however others have focused on cases that are less common in real world situations. The first uncommon case is when the heading angle is constant over time on the retina, despite rotation, e.g., when driving around a bend in a car and looking at a smudge on the windshield, such that one can see the world through the window in the visual periphery \cite{Stone1997HumanHE,Burlingham356758,Lappe1999PerceptionOS}.\footnote{This was likely the case tested by early psychophysics experiments due to the influence of the computer vision and robotics literatures. Cameras mounted to robots/cars fall under this case, as the camera only rotates when the robot or car changes direction.} The second uncommon case is when someone makes an eye movement to track a moving world object during ego-motion. This has typically been tested in past psychophysical studies with a straight path of travel and an object moving along an arc, such that its retinal velocity is constant, eliciting a constant velocity eye movement \cite{Royden1994,Banks1996}. In the first uncommon case, there is motion parallax, causing many singularities of retinal optic flow, each of whose position corresponds to a different depth plane \cite{Burlingham356758}. In this case, interestingly, if the path is circle (i.e., rotation and translation are constant over time) and the environment is a fronto-parallel plane, there is only one singularity of the optic acceleration field, i.e., the magnitude but not angles of the optic acceleration vectors depend on depth. And this singularity is aligned exactly with the heading. In this case, human observers are 3-10x more accurate in estimating their heading for displays that contain optic acceleration versus those that only contain optic flow (velocity) \cite{Burlingham356758}. No experiment has ever tested heading perception for a fixating observer while controlling for the presence of time-varying optic flow, so it's unclear whether the brain can estimate heading from the velocity field alone in this case. In this study, we tested the hypothesis that heading is computed by locating the singularity of optic acceleration \cite{Burlingham356758} and characterized the structure of the optic acceleration field for a fixating observer. We found that the singularity of optic acceleration was not an accurate estimate of heading, nor did it match human behavioral performance. In the process, we discovered that the structure of the optic acceleration and optic flow fields are more complex than we previously thought, containing two singularities with unique properties. 

The depth structure of the environment has a large influence on the structure of retinal motion. For an observer walking over a ground plane and fixating on a point off to the side on the ground, the optic flow field takes on a spiral structure \cite{Matthis2022} surrounding a singularity at the fixation point. This is because when you look down at the ground, each horizontal slice of the retinal image corresponds to a different depth, along a gradient. Therefore, each depth slice has slightly scaled translational optic flow, whereas the rotational optic flow doesn't depend on depth, and so the sum of these components creates a spiral pattern around the fixation point (motion parallax). We computed optic acceleration in this case and it has one singularity (always at the fixation point), which of course cannot provide information about heading. Human observers can perceive their heading accurately in this scenario \cite{Royden1994,van1993perception}, again suggesting that they don't use this estimator. 

For motion along a circular path with a time-invariant angle between gaze and heading, the singularity of optic acceleration doesn't depend on depth, whereas the flow field's does \cite{Burlingham356758}. This vastly simplifies the structure of the acceleration field for natural environments, for which there are many objects at different depths. On the other hand, the acceleration field for a fixating observer lacks this property. Indeed, the acceleration field changes substantially (i.e., 30º difference in singularity) for objects at depths nearby the fixation depth (Fig. 6). This means that the acceleration field is likely very complex (like the flow field) for a fixating observer in natural environments. 

For our simulations we assumed that the observer does not have vertical translational motion or non-yaw rotation, however generalizing those parameters to be non-zero does not affect our conclusion that the singularity of optic acceleration doesn't provide a good estimate of heading. On the contrary, our tests show that adding a vertical component of heading or non-yaw rotation generates acceleration fields that are at least as complex, and yields estimates of heading that are no better. 

A series of computer vision papers took a similar approach to us \cite{Bandyopadhyay1985PerceptionOR,Subbarao1989InterpretationOI,Barron1995MotionAS,Barron1996ComputationOT,Barron1996RecursiveEO,BARRON1996797}, introducing methods to estimate camera motion based on the singularity of optic acceleration. These methods assumed that camera motion is either constant or at most has constant acceleration and employed a Kalman filter to accumulate motion estimates over time, improving the accuracy and precision of the estimator. It is worth investigating how such methods achieved good accuracy, given the large degree of bias we observed in our own simulations. We speculate that it is because for the test videos they used, the camera's heading was yoked to its axis of rotation (like the assumption of fixating a smudge on the windshield while driving we made for Eq. \eqref{pnas_eq}), and rotational accelerations were small and noisy (in opposing directions, and hence their effects cancelled out over time), leading to low bias. 

Burlingham and Heeger (2020) showed that time-varying optic flow is necessary for accurate heading perception, but didn't rule out that instantaneous optic flow (velocity) is used. That is, the visual system could use optic acceleration alone, some combination of optic flow and optic acceleration, or other features in the time-varying retinal motion field. A number of previous methods use a combination of optic flow and acceleration to estimate heading \cite{Bandyopadhyay1985PerceptionOR,Wohn19863DMR,Subbarao1989InterpretationOI}. Specifically, these papers find that a solution for heading based on the spatial derivatives of optic flow is underconstrained in some cases, and that the spatio-temporal and/or temporal derivatives of optic flow can be introduced as additional constraints. Furthermore, it is known that a fixating observer provides an additional constraint on heading, because the rotation velocity of the eye and the translation velocity are related. \cite{Bandopadhay1986VisualNB} shows that a local estimator of heading based on optic flow is better constrained for a fixating observer than a non-fixating observer. Combining these ideas, Subbarao derived a solution for heading based on optic flow and optic acceleration for a fixating observer \cite{Subbarao1989InterpretationOI}. These early computer vision methods appear to be the best starting point for discovering how the human visual system computes heading. Future theoretical work should treat these methods (or variants of them) as models of visual perception and see which, if any, is consistent with human behavior and the neurophysiology of visual motion processing.

Because very few studies have investigated optic acceleration \cite{Edison2015HSGAAN,Edison_2017_CVPR_Workshops,Edison2019AutomatedVA}, there is currently no consensus on the best practices for estimating it. Barron and others \cite{Edison2015HSGAAN,Edison_2017_CVPR_Workshops,Edison2019AutomatedVA} estimated optic acceleration by taking a finite difference between two time-adjacent optic flow fields, a method known to produce biased solutions for discrete signals \cite{Farid_Simoncelli}. We developed and compared various methods for estimating optic acceleration, and propose a new method that is simple, fast, and accurate in the appendix of this paper (see Appendix).

\section{Code availability}
All code used to implement the retinal observer simulations and optic acceleration estimation methods is available at https://github.com/csb0/opticAcceleration.

%bibliography
%\bibliographystyle{unsrt}
%\bibliography{references}  %%% Uncomment this line and comment out the ``thebibliography'' section below to use the external .bib file (using bibtex) .

\clearpage 
\newpage 

%\onecolumn

\section{Appendix}

\subsection{Optic acceleration estimation method}
We developed an optic acceleration estimation method for use with videos and compared its performance to existing methods. Our method performed over an order of magnitude better than existing methods, as measured by angular error and endpoint error. We hope that our method will stimulate research on optic acceleration in computer vision and perceptual psychology.

The algorithm is simple and has three steps: (1) decimate (low-pass filter and downsample) the video if the resolution is too high, where 'too high' is determined by the local spatial variance of the image (i.e., producing image motion that is too fast in pixels/frame), (2) estimate optic flow fields for the entire image sequence using ref \cite{Farneback2000}, and (3) apply a 5-tap Farid \& Simoncelli temporal derivative filter \cite{Farid_Simoncelli} to the sequence of estimated optic flow fields to estimate optic acceleration.

\begin{algorithm}
\SetAlgoLined
\KwResult{Estimate of optic acceleration}
\If{size of each frame of the video is too large}{{decimate the video}}\;
 Estimate optic flow fields using Farneback's algorithm (2000) \cite{Farneback2000}\;
 Apply 5-tap Farid \& Simoncelli temporal derivative filter \cite{Farid_Simoncelli} to flow fields to compute acceleration fields\;
 \caption{Optic acceleration estimator "ffs"}
 \end{algorithm}
 
We compared this method to a number of existing and new estimation methods, detailed below.
 
\subsubsection{Test video generation}
We generated videos with which to test the accuracy of various optic acceleration estimation methods. The videos displayed either (1) a single moving object or (2) a planar environment while a camera traversed a circular path. Following Edison and Jiji \cite{Edison2015HSGAAN,Edison_2017_CVPR_Workshops,Edison2019AutomatedVA}, for (1) we used videos of a solitary moving circular disk. Unlike Edison and Jiji's disk, ours had a 1/f texture, which allowed for better estimation of optic flow within its edges. The disk moved downward with one of six acceleration magnitudes: 0.5, 1, 1.5, 2, 2.5, 3 pixels per frame$^2$ and each image was of size $800\times 800$ pixels. For (2) we simulated a circular path of travel toward an (initially) fronto-parallel plane composed of disks with 1/f textures (see \cite{Burlingham356758}'s Methods section for more detail). On a circular path, rotation and translation velocity are constant over time. Heading angle is thus constant and equivalent to the angle between the line-of-sight and the tangent of the circular path. In this case, the velocity of individual elements on the image plane depends on camera rotation, translation, and depth (see \cite{Burlingham356758}'s Supplemental Information for mathematical derivation). To generate a single video frame, we computed ground truth optic flow from these motion parameters and displaced each disk according to the optic flow at the nearest pixel location. We computed ground truth optic acceleration as the finite difference of consecutive ground truth optic flow fields.

\subsubsection{Comparison of optic acceleration estimation methods}
We evaluated the accuracy of various optic acceleration estimation methods on simulated videos with known optic acceleration. Following Edison and Jiji \cite{Edison2015HSGAAN,Edison_2017_CVPR_Workshops,Edison2019AutomatedVA}, we separately computed the mean endpoint or angular error between the ground truth and estimated optic acceleration vectors. For (1), we compared the endpoint or angle of the single ground truth acceleration vector with the spatial average of all estimated acceleration vectors within a disk's boundaries. For (2), we did the same, but computed the median angular or endpoint error across the entire acceleration field (i.e., for every disk in the display).   

Above is a table that compares the accuracy and efficiency of our optic acceleration estimation method with Edison's method and others we devised (Table 1). For this test, we used a circular path video with heading +9º and with rotation -6 º/s.
 
 \begin{table}
 \begin{tabularx}{\linewidth}
 {
 |>{\hsize=1.365\hsize}X
 |>{\hsize=0.3\hsize}X
 |>{\hsize=0.3\hsize}X
 |>{\hsize=0.3\hsize}X
 |>{\hsize=0.3\hsize}X
 |>{\hsize=0.3\hsize}X
 |}   
 \hline
 & mffd & mffs & hsfd & hsfs & ffs \\
 \hline
 Avg angular err (º) & 92.19 & 89.7 &90.02&92.7 &8.8 \\
 \hline
 Med angular err (º) & 94.18  &90.3   & 90.94  & 94.2  & 4.67  \\
 \hline
 Std angular err (º) & 6.88 & 6.88 & 6.66  & 6.69  & 1.95 \\
 \hline
 Avg endpt err (px) & 1.02  & 0.40  & 2.50  & 1.31  & 0.03 \\
  \hline
 Std endpt err (px) & 0.69  & 0.22  & 7.09  & 2.93  & 0.03 \\
 \hline
 Run time (s) & 3 & 3 &116 & 116 & 330 \\
  \hline
\end{tabularx}
\\
\caption{Error of various optic acceleration estimators when tested on a simulated navigation video containing global motion and with known ground truth optic flow and acceleration. Method names are shortened as follows. mffd: "Matlab" Farneback (2003) \& finite difference. mffs: "Matlab" Farneback (2003) \& Farid \& Simoncelli derivative filter. hsfd: Horn and Schunck \& finite difference. hsfs: Horn and Schunck \& \& Farid \& Simoncelli derivative filter. ffs, our method (Farneback (2020) and Farid \& Simoncelli derivative filter). }
 \end{table}
 
\subsubsection{Rationale}
Before developing this method, we first tried to find if there was an existing robust algorithm for estimating optic acceleration. Edison and colleagues \cite{Edison_2017_CVPR_Workshops} proposed and tested two methods — one that was simply a finite difference of optic flow over time, and another that was inspired by Horn \& Shunck's optic flow estimation method. They tested the accuracy of both methods on a video of small white moving disk that could accelerate and decelerate. The error they reported for this test was reasonable, but in our own extended tests we found that their method produced much larger errors (i.e., more than an order of magnitude greater) for white disks with larger accelerations than those they tested, when there were many moving objects, or when the object(s) had a non-uniform texture (e.g., a $1/f$ texture). We concluded that Edison's method \cite{Edison_2017_CVPR_Workshops} was insufficient for estimating optic acceleration for more realistic image motion.
 
Our proposed method relies on having prior estimates of optic flow. We tested a variety of different optic flow estimation algorithms, including Horn \& Schunck method \cite{Horn_Schunck}, Farneback's two-frame method from 2003 \cite{matlab_farneback} (built in as one option in the MATLAB function estimateFlow), Farneback's multi-frame, orientation tensor-based method from 2000 \cite{Farneback2000} and Lucas-Kanade's method \cite{lucas1981iterative}. We found that each of these algorithms produced comparably accurate estimates of optic acceleration for the video with a single moving uniform-colored disk, but only the Farneback (2000) based method performed well for a single moving 1/f textured disk and for the field of 1/f disks conveying an evolving retinal motion field corresponding to observer rotation and translation. Farneback (2000) takes multiple video frames as input, a spatiotemporal volume, and estimates motion as the orientation of a spacetime tensor using a parametric motion model. Using a multi-frame optic flow estimator is critical for generating accurate estimates of optic acceleration, because two-frame optic flow estimators are optimized for accurate estimation of the spatial structure of optic flow only and introduce substantial amounts of noise in the time course of optic flow at each pixel location. Subsequently taking the temporal derivative of optic flow amplifies this noise and because acceleration is usually a much smaller signal than velocity, the noise overtakes the signal and you get very poor estimates of acceleration.
 
To approximate the temporal derivative of the continuous, time-evolving optic flow that generated the discretely-sampled estimated optic flow, we employed Farid \& Simoncelli's 5-tap derivative filter \cite{Farid_Simoncelli}, designed for this purpose. This reduced errors substantially compared to a finite difference approximation of the temporal derivative (not shown in Table 1). 

Spatially downsampling / decimating the video beforehand was important to improve the efficacy and efficiency of the algorithm. Doing so makes the algorithm run much faster, without sacrificing accuracy, by reducing the size of each frame of the input video (the spatial derivative filters are the bottleneck, as they are slow for large inputs). And it ensures that there isn't image motion too large in pixels/frame, which no flow estimator can properly estimate. To ensure a fair comparison in Table 1, we downsampled the video inputs in the same way for each method.

\end{document}